\documentclass[conference,compsoc,onecolumn]{IEEEtran}
\ifCLASSOPTIONcompsoc
  \usepackage[nocompress]{cite}
\else
  \usepackage{cite}
\fi

\ifCLASSINFOpdf
\else
\fi
\usepackage{amsmath}

\usepackage{array}
\ifCLASSOPTIONcompsoc
\usepackage[caption=false,font=footnotesize,labelfont=sf,textfont=sf]{subfig}
\else
\usepackage[caption=false,font=footnotesize]{subfig}
\fi
\usepackage{url}
\usepackage{graphicx}
\newcommand{\HRule}{\rule{\linewidth}{0.25mm}}
\usepackage{algorithm,algpseudocode}
\algrenewcommand\algorithmicrequire{\textbf{Input:}}
\algrenewcommand\algorithmicensure{\textbf{Output:}}
\usepackage{amsmath,amsfonts,amssymb,amsthm}
\usepackage{xcolor}
\definecolor{ocre}{RGB}{0,173,239}

\usepackage{tikz}
\usepackage{pgfplots}
\pgfplotsset{compat=1.18}

\usetikzlibrary{arrows,matrix,positioning,fit}
\usetikzlibrary{patterns,decorations.pathreplacing}
\tikzset{%
  highlight/.style={rectangle,rounded corners,fill=ocre!50,draw=ocre!80,
     fill opacity=0.2,inner sep=0pt,text opacity=1}
}

\usepackage{graphicx} 
\usepackage{float}
\usepackage[all]{xy}
\usepackage{tikz-cd}
\usepackage[shortlabels]{enumitem}
\usepackage{dsfont}
\usepackage{tikz}
\usepackage{pbox}
\usetikzlibrary{backgrounds}
\pgfdeclarelayer{background2}
\pgfdeclarelayer{background}
\pgfdeclarelayer{foreground}
\pgfdeclarelayer{foreground2}
\pgfsetlayers{background2,background,main,foreground,foreground2}

\makeatletter
\renewcommand*\env@matrix[1][*\c@MaxMatrixCols c]{%
  \hskip -\arraycolsep
  \let\@ifnextchar\new@ifnextchar
  \array{#1}}
\makeatother

\usepackage[edges]{forest}
\usepackage{array}
\newcolumntype{C}[1]{>{\centering\arraybackslash}p{#1}}
\newcolumntype{L}[1]{>{\raggedright\arraybackslash}p{#1}}
\usetikzlibrary{arrows.meta,shadows}
\usetikzlibrary{trees,positioning,shapes,shadows,arrows}

\usepackage[hypertexnames=false]{hyperref}
\hypersetup{pdfborder=0 0 0}

\algnewcommand{\LineComment}[1]{\State \(\triangleright\) #1}

\usepackage{tablefootnote} % for table footnotes

\usepackage{lastpage} % for page out of
\usepackage{fancyhdr} % for page out of
\newlength{\gpgpuElemSep}
\newlength{\gpgpuElemSize}
\usepackage{listings}
\usepackage{caption}
\makeatletter
\let\old@lstKV@SwitchCases\lstKV@SwitchCases
\def\lstKV@SwitchCases#1#2#3{}
\makeatother
\usepackage{lstlinebgrd}
\makeatletter
\let\lstKV@SwitchCases\old@lstKV@SwitchCases

\lst@Key{numbers}{none}{%
    \def\lst@PlaceNumber{\lst@linebgrd}%
    \lstKV@SwitchCases{#1}%
    {none:\\%
     left:\def\lst@PlaceNumber{\llap{\normalfont
                \lst@numberstyle{\thelstnumber}\kern\lst@numbersep}\lst@linebgrd}\\%
     right:\def\lst@PlaceNumber{\rlap{\normalfont
                \kern\linewidth \kern\lst@numbersep
                \lst@numberstyle{\thelstnumber}}\lst@linebgrd}%
    }{\PackageError{Listings}{Numbers #1 unknown}\@ehc}}
\makeatother

\def\FunctionD(#1){2*(#1)-1}%
\def\FunctionM(#1,#2){((3*(#1)^2)*10^9)/(#2)}
\newcommand{\PreserveBackslash}[1]{\let\temp=\\#1\let\\=\temp}
\newcommand\hlightcolor[1]{\tikz[overlay, remember picture,baseline=-\the\dimexpr\fontdimen22\textfont2\relax]\node[rectangle,fill=magenta!50,rounded corners,fill opacity=0.2,draw=magenta!80,text opacity=1] {$#1$};} 
\usepackage{soul} %strikethrough font
\setstcolor{blue}
\usepackage{accents}

\definecolor{myblue4} {RGB}{66,129,196}
\tikzstyle{myline} = [very thick, draw=myblue4, fill=myblue4, shape=rectangle, inner sep=10pt, inner ysep=20pt]
\tikzstyle{mybox} = [very thick, draw=myblue4, fill=white, shape=rectangle, inner sep=10pt, inner ysep=20pt]

\tikzstyle{mybox1} = [very thick, draw=white, fill=white, shape=rectangle, inner sep=10pt, inner ysep=20pt]

\usepackage{adjustbox}
\usepackage{fourier-orns}
\usepackage{multirow}

\newtheorem{theorem}{Theorem}
\usepackage{tikz}
\usepackage{pgfplots}
\pgfplotsset{compat=1.18}
\usepackage{pgfplotstable}
\usepackage{ifthen}

\DeclareCaptionFormat{algor}{%
\hrulefill\par\offinterlineskip\vskip1pt%
\textbf{#1#2}#3\offinterlineskip\hrulefill}
\DeclareCaptionStyle{algori}{singlelinecheck=off,format=algor,labelsep=space}
\usepackage{bm}
\usepackage{setspace}

\usepackage{multicol}
\usepackage{pdflscape}

\usepackage{longtable}
\usepackage{stackengine}
\begin{document}
\pagenumbering{arabic}
\pagestyle{plain}

\title{Symbolic Algorithm for Solving SLAEs with Multi-Diagonal Coefficient Matrices}

\author{\IEEEauthorblockN{Milena Veneva}
\IEEEauthorblockA{Joint Institute for Nuclear Research, 6 Joliot-Curie St, Dubna, Moscow Region,
Russia,
141980,\\RIKEN Center for Computational Science, R-CCS, 7-1-26 Minatojima-minami-machi, Chuo-ku,\\ Kobe, Hyogo 650-0047, Japan,\\
Email: milena.p.veneva@gmail.com}}
% \and
% \IEEEauthorblockN{Alexander Ayriyan}
% \IEEEauthorblockA{
% Email: }
% \and
% \IEEEauthorblockN{Toshiyuki Imamura}
% \IEEEauthorblockA{
% Email: }}

% make the title area
\maketitle
\thispagestyle{plain}

\section*{Abstract}
This paper presents a generalised symbolic algorithm for solving systems of linear algebraic equations with multi-diagonal coefficient matrices. The algorithm is given in a pseudocode. A theorem which gives the condition for correctness of the algorithm is formulated and proven. Formula for the complexity of the multi-diagonal numerical algorithm is obtained.

\section{Introduction} Systems of linear algebraic equations (SLAEs) with multi-diagonal coefficient matrices may arise after many different scientific and engineering problems, as well as problems of the computational
linear algebra where finding the solution of an SLAE is considered to be one of the most important problems. For instance,
the resultant SLAE after discretization of partial differential equations (PDEs), using finite difference methods (FDM) or
finite element methods (FEM) has a banded coefficient matrix. The methods for solving such SLAEs known in the literature
usually require the matrix to possess special characteristics so as the method to be numerically correct and stable, e.\,g.\,diagonal dominance, positive
definiteness, etc.
%\,which are not always feasible. 
Another possible approach which ensures numerically correct formulae without adding special additional requirements or using pivoting is the symbolic algorithms.

By definition, a band SLAE is an SLAE with band coefficient matrix. The lower band width $p$ is the number of sub-diagonals, the upper band width $q$ is the number of super-diagonals, and the band width of the matrix is defined as $p+q+1$ (we should add $1$ because of the main diagonal), that is, the total number of non-zero diagonals in the matrix~\cite{Spiteri_2007}. Here, we are going to focus on SLAEs with matrices for which $p=q=M$. The author of~\cite{Christov_1994} presents a generalised numerical algorithm for solving multi-diagonal SLAEs with pivoting (implemented in \texttt{Fortran}) in the case $p\neq q$, and has applied it for solving boundary value problems discretized by finite difference approximations. 

%\subsection{Symbolic algorithms}

A whole branch of symbolic algorithms for solving systems of linear algebraic equations with different coefficient matrices exists in the literature. For instance, in \cite{El-Mikkawy_2012}~the author considers a tridiagonal matrix and a symbolic version of the Thomas method~\cite{Thomas_1949,Higham_2002} is formulated. The authors of~\cite{Karawia_2013a}
build an algorithm in the case of a general bordered tridiagonal SLAE, while in~\cite{Atlan_2015} the coefficient matrix taken into consideration is a general
opposite-bordered tridiagonal one. 

A pentadiagonal coefficient matrix is of
interest in~\cite{Askar_2015}, while a cyclic pentadiagonal coefficient matrix is considered in~\cite{Jia_2012}. The latter algorithm can be applied to periodic
tridiagonal and periodic pentadiagonal SLAE either by setting the corresponding
matrix terms to be zero.

In \cite{Karawia_2013b} a symbolic method for the case of a cyclic heptadiagonal SLAEs is presented. 

What is common for all these symbolic algorithms, is that they are implemented using Computer Algebra Systems (CASs) such as \texttt{Maple}~\cite{maple}, \texttt{Mathematica}~\cite{mathematica}, and \texttt{Matlab}~\cite{matlab}.

Finally, \cite{Veneva_2018b} presents a symbolic method for the case of a pure heptadiagonal SLAE.

%A performance analysis of effective methods (both numerical and symbolic)
A performance analysis of symbolic methods (and numerical as well)
for solving band
matrix SLAEs (with three and five diagonals) being implemented in \texttt{C++} and run on modern (as of
2018) computer systems is made in~\cite{Veneva_2019a}. Different strategies (symbolic included) for solving
band matrix SLAEs (with three and five diagonals) are explored in~\cite{Veneva_2018a}. A performance
analysis of effective symbolic algorithms for solving band matrix SLAEs with coefficient matrices
with three, five and seven diagonals being implemented in both \texttt{C++} and \texttt{Python} and run on modern
(as of 2018) computer systems is made in~\cite{Veneva_2019b}. 

Having in mind all these introductory notes, it is clear that a generalised multi-diagonal symbolic algorithm is the novelty that addresses the need of a direct method which solves multi-diagonal systems of linear algebraic equations without putting any requirements for the characteristics of the coefficient matrix. Thus, the aim of this paper, which is a logical
continuation of~\cite{Veneva_2018b, Veneva_2019a, Veneva_2018a, Veneva_2019b}, is to present such a generalised symbolic algorithm for solving SLAEs with multi-diagonal coefficient matrices. The symbolic algorithms investigated in~\cite{Veneva_2018b, Veneva_2019a, Veneva_2018a, Veneva_2019b} are actually particular cases of the generalised multi-diagonal symbolic method when $p = q = M = 1, 2$, and $3$.

The layout of the paper is as follows: in the next section, we outline the multi-diagonal numerical algorithm, and 
introduce the multi-diagonal symbolic algorithm in pseudocode. Afterwards, we make some correctness remarks for the symbolic method, and present a generalised formula for the complexity of the multi-diagonal numerical algorithm. Finally, some conclusions are drawn.

The novelties of this work are as follows: suggested multi-diagonal symbolic algorithm for solving SLAEs, formulation and proof of a correctness theorem, and an additionally obtained formula for the complexity of the multi-diagonal numerical method.

\section{Multi-diagonal symbolic algorithm}

Let us consider an SLAE $Ax = y$, where 
$A=\textrm{diag}(\mathbf{b^{0}}, \mathbf{b^{1}}, \mathbf{b^{2}}, \ldots, \mathbf{b^{M}}, \mathbf{b^{M+1}},
\mathbf{b^{M+2}}, \mathbf{b^{M+3}}, \ldots, \mathbf{b^{2\times M}})$, $A$ is a real $N\times N$ multi-diagonal matrix with $M$ sub-diagonals, and $M$ super-diagonals, and $2\times M + 1 < N$, that is, the number of diagonals should be smaller than the number of equations within the SLAE; $x$ and $y$ are real column vectors with $N$ elements:
\begin{equation}
\begin{bmatrix}
b_{0}^{M} & b_{0}^{M+1} & b_{0}^{M+2} & b_{0}^{M+3} & \dots & b_{0}^{2M} & 0 & \dots  & \dots & \dots & \dots & 0 \\
b_{1}^{M-1} & b_{1}^{M} & b_{1}^{M+1} & b_{1}^{M+2} & \dots & b_{1}^{2M-1} & b_{1}^{2M} & 0 & \dots & \dots & \dots & 0 \\
b_{2}^{M-2} & b_{2}^{M-1} & b_{2}^{M} & b_{2}^{M+1} & \dots & b_{2}^{2M-2} & b_{2}^{2M-1} & b_{2}^{2M} & 0 & \dots & \dots & 0 \\
b_{3}^{M-3} & b_{3}^{M-2} & b_{3}^{M-1} & b_{3}^{M} & \dots & b_{3}^{2M-3} & b_{3}^{2M-2} & b_{3}^{2M-1} & b_{3}^{2M} & 0 & \dots & 0 \\
b_{4}^{M-4} & b_{4}^{M-3} & b_{4}^{M-2} & b_{4}^{M-1} & \dots & b_{4}^{2M-4} & b_{4}^{2M-3} & b_{4}^{2M-2} & b_{4}^{2M-1} & b_{4}^{2M} & \dots & 0 \\
\vdots & \ddots & \ddots & \ddots & \ddots & \ddots & \ddots & \ddots & \ddots & \ddots & \ddots & \vdots \\
0 & \dots & 0 & b_{N-4}^{0} & b_{N-4}^{1} & \dots & \dots & b_{N-4}^{M-1} & b_{N-4}^{M} & b_{N-4}^{M+1} & b_{N-4}^{M+2} & b_{N-4}^{M+3}\\
0 & \dots & \dots & 0 & b_{N-3}^{0} & b_{N-3}^{1} & \dots & \dots & b_{N-3}^{M-1} & b_{N-3}^{M} & b_{N-3}^{M+1} & b_{N-3}^{M+2}\\
0 & \dots & \dots & \dots & 0 & b_{N-2}^{0} & b_{N-2}^{1} & \dots & \dots & b_{N-2}^{M-1} & b_{N-2}^{M} & b_{N-2}^{M+1} \\
0 & \dots & \dots & \dots & \dots & 0 & b_{N-1}^{0} & b_{N-1}^{1} & \dots & \dots & b_{N-1}^{M-1} & b_{N-1}^{M}
\end{bmatrix}
\begin{bmatrix}
x_{0} \\
x_{1} \\
x_{2} \\
x_{3} \\
x_{4} \\
\vdots \\
x_{N-4} \\
x_{N-3} \\
x_{N-2} \\
x_{N-1}
\end{bmatrix}
=
\begin{bmatrix}
y_{0} \\
y_{1} \\
y_{2} \\
y_{3} \\
y_{4} \\
\vdots \\
y_{N-4} \\
y_{N-3} \\
y_{N-2} \\
y_{N-1}
\end{bmatrix}.
\end{equation}

The multi-diagonal numerical solver which we are going to formulate below is a generalization of the Thomas method for multi-diagonal SLAEs. The algorithm is based on LU decomposition, and requires forward reduction for reducing the initial matrix into a lower triangular one: 

\begin{equation*}
\begin{aligned}[t]
&\mu_{0} = b_{0}^{M} \\
&\alpha_{0}^{M+1} = \frac{b_{0}^{M+1}}{\mu_{0}} \\
&\alpha_{0}^{M+2} = \frac{b_{0}^{M+2}}{\mu_{0}} \\
&\dots \\
&\alpha_{0}^{2M} = \frac{b_{0}^{2M}}{\mu_{0}} \\
&z_{0} = \frac{y_{0}}{\mu_{0}} \\
\end{aligned}
%\qquad
\hspace{6em}
\vrule{}
\hspace{6em}
\begin{aligned}[t]
&\alpha_{1}^{j} = 0,\quad i=1,2\ldots,M-1 \\
&\alpha_{1}^{M} = b_{1}^{M-1}\\
&\mu_{1} = b_{1}^{M} - \alpha_{0}^{M+1} \times \alpha_{1}^{M}\\
&\alpha_{1}^{M+1} = \frac{b_{1}^{M+1} - \alpha_{0}^{M+2}\times \alpha_{1}^{M}}{\mu_{1}} \\
&\alpha_{1}^{M+2} = \frac{b_{1}^{M+2} - \alpha_{0}^{M+3}\times \alpha_{1}^{M}}{\mu_{1}} \\
& \dots \\
&\alpha_{1}^{2M} = \frac{b_{1}^{2M}}{\mu_{1}} \\ 
&z_{1} = \frac{y_{1} - z_{0}\times \alpha_{1}^{M}}{\mu_{1}} \\
\end{aligned}
\end{equation*}
\noindent\rule{\textwidth}{0.4pt}
\begin{equation*}
\hspace{-6em}\begin{aligned}[t]
&\textrm{For }i = 2,3,\dots,M-1:\\
&\textrm{counter} = M - i\\
&\alpha_{i}^{k-\textrm{counter}} = b_{i}^{k-1},\quad 
k= M, M-1,\ldots,1,\quad k -\textrm{counter} \geq 1 \\
&\alpha_{i}^{k-\textrm{counter}} = \alpha_{i}^{k-\textrm{counter}} - 
\alpha_{0}^{M + k - \textrm{counter}-1}\times\alpha_{i}^{1} - 
\alpha_{1}^{M + k - \textrm{counter} - 2}\times\alpha_{i}^{2} -
\dots - \alpha_{k - \textrm{counter}-2}^{M+1}\times\alpha_{i}^{k - \textrm{counter} -1},\quad\\
&\qquad k =2,3,\ldots, M\\
&\mu_{i} = b_{i}^{M} - \alpha_{0}^{M + i}\times\alpha_{i}^{1} - \alpha_{1}^{M + i-1}\times\alpha_{i}^{2}-\dots-\alpha_{i-1}^{M+1}\times\alpha_{i}^{i}\\
&\alpha_{i}^{M+1} = \frac{b_{i}^{M+1} - \alpha_{0}^{M+i+1}\times \alpha_{i}^{1}-
\alpha_{1}^{M+i}\times \alpha_{i}^{2}-\dots-\alpha_{i-1}^{M+2}\times \alpha_{i}^{i}}{\mu_{i}} \\
& \textrm{num\_sub} = min(i, M - 2) \quad\textrm{number of subtractions for } \alpha_{i}^{M+2}\\
% &\alpha_{i}^{M+2} = \dfrac{1}{\mu_{i}}\left(b_{i}^{M+2} - \alpha_{i-\textrm{num\_sub}}^{M+2+\textrm{num\_sub}}\times \alpha_{i}^{i-\textrm{num\_sub}+1}\right.\left.-
% \alpha_{i-\textrm{num\_sub}+1}^{M+2+\textrm{num\_sub} - 1}\times \alpha_{i}^{i-\textrm{num\_sub}+2}-\dots-\alpha_{i-1}^{M+3}\times \alpha_{i}^{i}\right) \\
% & \dots \\
% & \textrm{num\_sub} = min(i, M - 1 - (k - 1)) \quad\textrm{number of subtractions for } \alpha_{i}^{M+k}\\
% & \dots \\
% &\alpha_{i}^{2M} = \frac{b_{i}^{2M}}{\mu_{i}} \\ 
\end{aligned}
\end{equation*}
%%%%% ottuk
\begin{equation*}
\hspace{-6em}\begin{aligned}[t]
%&\textrm{For }i = 2,3,\dots,M-1:\\
%&\textrm{counter} = M - i\\
%&\alpha_{i}^{k-\textrm{counter}} = b_{i}^{k-1},\quad 
%k= M, M-1,\ldots,1,\quad k -\textrm{counter} \geq 1 \\
%&\alpha_{i}^{k-\textrm{counter}} = \alpha_{i}^{k-\textrm{counter}} - 
%\alpha_{0}^{M + k - \textrm{counter}-1}\times\alpha_{i}^{1} - 
%\alpha_{1}^{M + k - \textrm{counter} - 2}\times\alpha_{i}^{2} -
% \dots - \alpha_{k - \textrm{counter}-2}^{M+1}\times\alpha_{i}^{k - \textrm{counter} -1},\quad\\
% &\qquad k =2,3,\ldots, M\\
% &\mu_{i} = b_{i}^{M} - \alpha_{0}^{M + i}\times\alpha_{i}^{1} - \alpha_{1}^{M + i-1}\times\alpha_{i}^{2}-\dots-\alpha_{i-1}^{M+1}\times\alpha_{i}^{i}\\
% &\alpha_{i}^{M+1} = \frac{b_{i}^{M+1} - \alpha_{0}^{M+i+1}\times \alpha_{i}^{1}-
% \alpha_{1}^{M+i}\times \alpha_{i}^{2}-\dots-\alpha_{i-1}^{M+2}\times \alpha_{i}^{i}}{\mu_{i}} \\
% & \textrm{num\_sub} = min(i, M - 2) \quad\textrm{number of subtractions for } \alpha_{i}^{M+2}\\
&\alpha_{i}^{M+2} = \dfrac{1}{\mu_{i}}\left(b_{i}^{M+2} - \alpha_{i-\textrm{num\_sub}}^{M+2+\textrm{num\_sub}}\times \alpha_{i}^{i-\textrm{num\_sub}+1}\right.\left.-
\alpha_{i-\textrm{num\_sub}+1}^{M+2+\textrm{num\_sub} - 1}\times \alpha_{i}^{i-\textrm{num\_sub}+2}-\dots-\alpha_{i-1}^{M+3}\times \alpha_{i}^{i}\right) \\
& \dots \\
& \textrm{num\_sub} = min(i, M - 1 - (k - 1)) \quad\textrm{number of subtractions for } \alpha_{i}^{M+k}\\
& \dots \\
&\alpha_{i}^{2M} = \frac{b_{i}^{2M}}{\mu_{i}} \\ 
\end{aligned}
\end{equation*}
\begin{equation*}
\hspace{-6em}\begin{aligned}[t]
&\hspace{-22em}z_{i} = \frac{y_{i} - z_{0}\times \alpha_{i}^{1}- z_{1}\times \alpha_{i}^{2}-\dots-z_{i-1}\times \alpha_{i}^{i}}{\mu_{i}} \\
\end{aligned}
\end{equation*}
\noindent\rule{\textwidth}{0.4pt}
\begin{equation*}
\begin{aligned}[t]
&\textrm{For }i = M,M+1,\dots,N-1:\\
&\alpha_{i}^{j} = b_{i}^{j-1},\quad j=1,2\ldots,M-1 \\
&\alpha_{i}^{k} = \alpha_{i}^{k} - 
\alpha_{i-M}^{M+k-1}\times \alpha_{i}^{1} - \alpha_{i-M+1}^{M+k-2}\times \alpha_{i}^{2} - \dots -\alpha_{i-M+\textrm{iter}}^{M+k-1-\textrm{iter}}\times \alpha_{i}^{1+\textrm{iter}}, \\
&\quad\quad k=2,3,\ldots,M, \quad \textrm{iter}=0,1,\ldots,k-2,\quad i=M,M+1\ldots,N-1\\
&\mu_{i} = b_{i}^{M} - \alpha_{i-M}^{2M}\times\alpha_{i}^{1} - \alpha_{i-M+1}^{2M-1}\times\alpha_{i}^{2}-\dots-\alpha_{i-1}^{M+1}\times\alpha_{i}^{M}
,\qquad\,\, i=M,M+1,\ldots,N-1 \\
&\alpha_{i}^{M+1} = \frac{b_{i}^{M+1}-\alpha_{i-1}^{M+2}\times\alpha_{i}^{M}-\alpha_{i-2}^{M+3}\times\alpha_{i}^{M-1}-\alpha_{i-3}^{M+4}\times\alpha_{i}^{M-2}-\dots-\alpha_{i-M+1}^{2M}\times \alpha_{i}^{2}}{\mu_{i}},\qquad\,\, i=M,M+1,\ldots,N-2 \\
&\alpha_{i}^{M+2} = \frac{b_{i}^{M+2}-\alpha_{i-1}^{M+3}\times\alpha_{i}^{M}-\alpha_{i-2}^{M+4}\times\alpha_{i}^{M-1}-\alpha_{i-3}^{M+5}\times\alpha_{i}^{M-2}-\dots-\alpha_{i-M+2}^{2M}\times \alpha_{i}^{3}}{\mu_{i}},\qquad\,\, i=M,M+1,\ldots,N-3 \\
&\dots \\
&\alpha_{i}^{2M} = \frac{b_{i}^{2M}}{\mu_{i}},\qquad\,\, i=M,M+1,\ldots,N-1-M \\
&z_{i} = \frac{y_{i} - z_{i-M}\times\alpha_{i}^{1} - z_{i-M+1}\times\alpha_{i}^{2}- \dots - z_{i-1}\times \alpha_{i}^{M}}{\mu_{i}},\qquad\,\, i=M,M+1,\ldots,N-1 \\
\end{aligned}
\end{equation*}

and a backward substitution for finding the unknowns $x$ in a reverse order:

\begin{equation*}
\begin{aligned}
x_{N-1} &= z_{N-1} \\
x_{N-k} &= z_{N-k} - \alpha_{N-k}^{M+1}\times x_{N-k+1} - \dots -\alpha_{N-k}^{M+1+k-2}\times x_{N-1}, \quad k=2,3,\ldots, M\\
x_{i} &= z_{i} - \alpha_{i}^{M+1}\times x_{i+1} - \alpha_{i}^{M+2}\times x_{i+2} - \alpha_{i}^{M+3}\times x_{i+3} - \dots - \alpha_{i}^{2M}\times x_{i+M},\\
&\quad i=N-(M+1),N-(M+2),\ldots,0. 
\end{aligned}    
\end{equation*} 

In order to
cope with the stability issue of the Thomas method in the case of non-diagonally dominant matrices, in the case of a zero (or numerically zero) quotient of two subsequent leading
principal minors within the symbolic method a symbolic variable is assigned instead and the calculations are
continued. At the end of the algorithm, this symbolic variable is substituted with zero.
The same approach is suggested in~\cite{El-Mikkawy_2012}.

The full multi-diagonal symbolic method in pseudocode is given in Algorithm~\ref{alg:multi_symb}. There, $\varepsilon$ plays the role of a numerical zero, and was set to $1.0\mathrm{e}{-20}$ in our code.
%\newpage
%\\\HRule\\[-0.1cm]
%\textbf{Algorithm 4.} Heptadiagonal symbolic algorithm for solving an SLAE $Ax = y$.
%\\[-0.5em]
%\HRule\\[-0.1cm]
\begin{center}
\captionof{algorithm}{Multi-diagonal symbolic algorithm for solving an SLAE $Ax = y$.}\label{alg:multi_symb}
\begin{multicols}{2}
\begin{algorithmic}[1]\setstretch{1.8}
\Require{$N, \mathbf{b^{0}}, \mathbf{b^{1}}, \dots, \mathbf{b^{M}}, \mathbf{b^{M+1}, \dots, \mathbf{b^{2M}}},  \mathbf{y}, \varepsilon$}
\Ensure{$\mathbf{x}$}
\If{$det(A)==0$}\State{exit}\EndIf
\State{bool flag = False}
\State{$\mu_{0}=b_{0}^{M}$} \Comment{Step 1.(0)}
\If{$|\mu_{0}|<\varepsilon$} \State{$\mu_{0}=$ symb; flag = True}\EndIf
\For{$k=\overline{M+1,\ldots, 2M}$}
\State{$\alpha_{0}^{k}=\dfrac{b_{0}^{k}}{\mu_{0}}$}
\EndFor
\State{$z_{0}=\dfrac{y_{0}}{\mu_{0}}$}
% 1st
\For{$k=\overline{1,2,\ldots, M}$}
\State{$\alpha_{1}^{k}=b_{1}^{k-1}$}
\EndFor
\Comment{(1)}
\State{$\mu_{1}=b_{1}^{M}-\alpha_{0}^{M+1}\times\alpha_{1}^{M}$} 
\If{!flag}\If{$|\mu_{1}|<\varepsilon$}\State{$\mu_{1}=$ symb; flag = True}\EndIf\EndIf
\For{$k=\overline{M+1,\ldots, 2M}$}
\State{$\alpha_{1}^{k}=b_{1}^{k}$}
\If{$M > 1$ and $k < 2M$}
\State{$\alpha_{1}^{k}=\alpha_{1}^{k} - \alpha_{0}^{k+1}\times\alpha_{1}^{M}$}
\EndIf
\State{$\alpha_{1}^{k}=\dfrac{\alpha_{1}^{k}}{\mu_{1}}$}
\EndFor
\State{$z_{1}=\dfrac{y_{1}- z_{0}\times\alpha_{1}^{M}}{\mu_{1}}$}
% i-th
\For{$i=\overline{2,\ldots, N-1}$}
\State{counter = 0}
\Comment{number of non-zero helping $\alpha_{1}^{k}$,\\}
\Comment{where $k=\overline{1,2,\ldots, M}$} 
\If{$i < M$}\State{$\textrm{counter} = M - i$}\EndIf
\For{$k=\overline{1,\ldots, M}$}
\If{$k-\textrm{counter}\geq 1$}
\State{$\alpha_{i}^{k-\textrm{counter}}=b_{i}^{k-1}$}
\EndIf
\EndFor\\
\Comment{above we shift the non-zero $\alpha_{i}^{j}, j\leq M$ in order\\} 
\Comment{to have them in the interval $j\in[0; ...]$}
\For{$k=\overline{M+1,\ldots, 2M}$}
\State{$\alpha_{i}^{k}=b_{i}^{k}$}
\EndFor
\State{$\mu_{i}=b_{i}^{M}$} 
\State{$z_{i}=y_{i}$} 
\State{$\textrm{iter} = 0$}\\
\Comment{number of iterations for $\alpha_{i}^{k}$, where $k\leq M$} 
\State{$\textrm{coeff} = 0$}\\
\Comment{the biggest distance between the lower coeff of}\\
\Comment{$\alpha_{i}^{k}$ and $\alpha_{\textrm{coeff} + \textrm{iter}}^{M + ...}$}
\If{$i >= M$}\State{$\textrm{coeff} = i - M$}\EndIf
\For{$k=\overline{2,\ldots, M}$}
\State{$\textrm{iter} = 0$}
\For{$l=\overline{2,\ldots, k-\textrm{counter}}$}
%\If{$k - 1 - \textrm{iter} \leq 0$}\State{break}\EndIf
\State{$\alpha_{i}^{k-\textrm{counter}} = \alpha_{i}^{k-\textrm{counter}}$}
\State{\hspace{4em}$-\alpha_{\textrm{coeff}+\textrm{iter}}^{M+k-1-\textrm{counter}-\textrm{iter}}~\times~\alpha_{i}^{1 + \textrm{iter}}$}
\State{$\textrm{iter} = 1 + \textrm{iter}$}
\EndFor
\EndFor
\State{$\textrm{iter} = 0$}\\
\Comment{number of iterations for $\mu_{i}$ and $z_{i}$}
\If{$i < M$}\State{$\textrm{mu\_max\_iter} = i - 1$}
\Else\State{$\textrm{mu\_max\_iter} = M - 1$}\EndIf
\For{$\textrm{iter}=\overline{0,1,\ldots,\textrm{mu\_max\_iter}}$}
\Comment{$\mu_{i}, z_{i}$}
%\If{$i < M$}\If{$\textrm{iter} \geq i$}\State{break}\EndIf
%\ElsIf{$\textrm{iter} \geq M$}
%\State{break}
%\EndIf
\State{$\mu_{i} = \mu_{i} - \alpha_{\textrm{coeff}+\textrm{iter}}^{2M - \textrm{counter} - \textrm{iter}}\times\alpha_{i}^{1 + \textrm{iter}}$}
\State{$z_{i} = z_{i} - z_{\textrm{coeff}+\textrm{iter}}\times\alpha_{i}^{1 + \textrm{iter}}$}
%\State{$\textrm{iter} = \textrm{iter} + 1$}
\EndFor
\If{!flag}\If{$|\mu_{i}|<\varepsilon$}\State{$\mu_{i}=$ symb; flag = True}\EndIf\EndIf
\State{$z_{i}=\dfrac{z_{i}}{\mu_{i}}$}
% follows alpha
\State{$\textrm{iter} = 0$}
\Comment{number of iterations for a}\\
\Comment{particular $\alpha_{i}^{m}, m \geq M + 1$}
\State{$\textrm{alpha\_counter} = 0$} 
\Comment{number of $\alpha_{i}^{m}, m \geq M + 1$} 
\For{$m=\overline{0,1,\ldots, M - 1}$}
\State{$\textrm{num\_sub}[m] = 0$}\\
\Comment{number of subtractions in $\alpha_{i}^{m}$}
\EndFor
\For{$m=\overline{M+1,\ldots, 2M - 1}$}
\State{$\textrm{m\_index} = m - M - 1$}\\
\Comment{shift the index from $0$ to $M - 2$}
% \If{$M > 1$}\State{$\textrm{num\_sub}[\textrm{m\_index}] = 1$}\EndIf
\If{$i \leq M - 1$} 
%and $\textrm{num\_sub}[\textrm{m\_index}] <$\\\hspace{4em} $M - 1 - \textrm{alpha\_counter}$}
\State{$\textrm{num\_sub}[\textrm{m\_index}] =$\\\hspace{6em}$min(i, M - 1 - \textrm{alpha\_counter})$}
%\EndIf
%\If{$i \geq M$}
\Else\State{$\textrm{num\_sub}[\textrm{m\_index}] =$\\\hspace{6em}$M - 1 - \textrm{alpha\_counter}$}\EndIf
\State{$\textrm{iter} = 0$}
\For{$k=\overline{0,1,\ldots, \textrm{num\_sub}[\textrm{m\_index}] - 1}$}
\State{$\textrm{coeff} = i - \textrm{num\_sub}[\textrm{m\_index}] + \textrm{iter}$}
% \If{$i \geq M$}\State{$\textrm{coeff} = i - \textrm{num\_sub}[\textrm{m\_index}] + \textrm{iter}$}
% \Else
% \State{$\textrm{coeff} = i - \textrm{num\_sub}[\textrm{m\_index}] + \textrm{iter}$}
% \EndIf
\State{$\textrm{coeff\_1} = 0$}
\If{$i \geq M$}\State{$\textrm{coeff\_1} =(M - \textrm{num\_sub}[\textrm{m\_index}]$\\\hspace{12em}$+ \textrm{iter}) \% M + 1$}
\Else
\State{$\textrm{coeff\_1} = i - \textrm{num\_sub}[\textrm{m\_index}]$}
\State{\hspace{5em}$+ \textrm{iter} + 1$}
\EndIf\\
\Comment{the helping $\alpha_{i}^{\textrm{coeff\_1}}$ are with upper index\\}
\Comment{ up to $M$, therefore we need to find the\\}
\Comment{module($M$)}
\State{$\alpha_{i}^{m} = \alpha_{i}^{m}$}
\State{$\hspace{1.6em}- \alpha_{\textrm{coeff}}^{m + \textrm{num\_sub}[\textrm{m\_index}] - \textrm{iter}}\times\alpha_{i}^{\textrm{coeff\_1}}$}
\State{$\textrm{iter} = \textrm{iter} + 1$}
\EndFor
\State{$\textrm{alpha\_counter} = \textrm{alpha\_counter} + 1$}
\For{$k=\overline{M + 1,\ldots, 2M}$}
\State{$\alpha_{i}^{k} = \dfrac{\alpha_{i}^{k}}{\mu_{i}}$}
\EndFor
\EndFor
\EndFor
\State{$x_{N-1}=z_{N-1}$} \Comment{Step 2. Solution}
\For{$\overline{i=N-2,\ldots 0}$}
\State{$x_{i}=z_{i}$}
\State{$\textrm{iter} = 0$}
\For{$\overline{k=0,\ldots M - 1}$}
\If{$i + \textrm{iter} > n - 2$}\State{break}\EndIf
\State{$x_{i}=x_{i} - \alpha_{i}^{M + 1 + k}\times x_{i + 1 + k}$}
\State{$\textrm{iter} = \textrm{iter} + 1$}
\EndFor
\EndFor
\State{Cancel the common factors in the numerators and denominators of
$\mathbf{x}$, making them coprime. Substitute $\textrm{symb}:=0$ in
$\mathbf{x}$ and simplify.}
\end{algorithmic}
\end{multicols}
\vspace{-0.5em}
\HRule
\end{center}

\textbf{Remark:} If any $\mu_{i}$ expression has been evaluated to be zero or
numerically zero, then it is assigned to be a symbolic variable. We cannot
compare any of the next $\mu_{i}$ expressions with $\varepsilon$, because any
further $\mu_{i}$ is going to be a symbolic expression. To that reason, we use a
boolean flag which tells us if any previous $\mu_{i}$ is a symbolic expression. In
that case, comparison with $\varepsilon$ is not conducted as being not needed.

\section{Justification of the algorithm}

Let us make some observations on the correctness of the proposed algorithm. In case the algorithm assigns $\mu_{i}$ for any $i=\overline{0, N-1}$ to be equal to a symbolic
variable (in case $\mu_{i}$ is zero or numerically zero), this ensures correctness of the
formulae for computing the solution of the considered SLAE (because we are not dividing by (numerical) zero). 
However, this does
not add any additional requirements to the coefficient matrix so as to keep the algorithm stable.
\begin{theorem}
The only requirement to the coefficient matrix of a multi-diagonal SLAE so as the multi-diagonal symbolic algorithm to be
correct is nonsingularity.
\end{theorem}
\begin{proof}
As a direct consequence of the transformations done so as the matrix $A$ to be
factorized and then the downwards sweep to be conducted, it follows that the
determinant of the matrix $A$ in the terms of the introduced notation is:
\begin{equation}
\label{eq_1}
\textrm{det}(A) = \prod_{i=0}^{N-1}\mu_{i}|_{\textrm{symb}=0},
\end{equation}
because the determinant of an upper triangular matrix is equal to the product of all its diagonal elements~\cite{Kaye_1998}. (This formula could be used so as the nonsingularity of the coefficient matrix
to be checked.)
If $\mu_{i}$ for any $i$ is assigned to be equal to a symbolic variable, then
it is going to appear in both the numerator and the denominator of the
expression for the determinant and so it can be cancelled:
\begin{equation}
\begin{aligned}
\label{eq_2}
\textrm{det}(A)
&= \mu_{0}\,\mu_{1}\,\mu_{2}\ldots\mu_{N-2}\,\mu_{N-1} = \\
&= M_{0}\,\frac{M_{1}}{\mu_{0}}\,\frac{M_{2}}{\mu_{0}\,\mu_{1}}\ldots
\frac{M_{N-2}}{\mu_{0}\,\mu_{1}\ldots\mu_{N-3}}\,\frac{M_{N-1}}
{\mu_{0}\,\mu_{1}\ldots\mu_{N-2}} = \\
& = \frac{\prod_{i=0}^{N-1}M_{i}}{\mu_{0}^{N-1}\,\mu_{1}^{N-2}\,\mu_{2}^{N-3}
\ldots\mu_{N-3}^{2}\,\mu_{N-2}^{1}} = \\
& = \frac{\prod_{i=0}^{N-1}M_{i}}{M_{0}^{N-1}\,\frac{M_{1}^{N-2}}{\mu_{0}^{N-2}}\,
\frac{M_{2}^{N-3}}{\mu_{0}^{N-3}\,\mu_{1}^{N-3}}\ldots\frac{M_{N-3}^{2}}
{\mu_{0}^{2}\,\mu_{1}^{2}\ldots\mu_{N-4}^{2}}\,\frac{M_{N-2}^{1}}{\mu_{0}^{1}\,
\mu_{1}^{1}\ldots\mu_{N-3}^{1}}} = \\
& = \frac{\prod_{i=0}^{N-1}M_{i}}{M_{0}^{N-1}\,\frac{M_{1}^{N-2}}{\mu_{0}^{N-2}}\,
\frac{M_{2}^{N-3}}{\mu_{0}^{N-3}\,\left(\frac{M_{1}}{\mu_{0}}\right)^{N-3}}\ldots
\frac{M_{N-3}^{2}}
{\mu_{0}^{2}\,\left(\frac{M_{1}}{\mu_{0}}\right)^{2}\ldots
\left(\frac{M_{N-4}}{\mu_{N-3}}\right)^{2}}\,\frac{M_{N-2}^{1}}{\mu_{0}^{1}\,
\left(\frac{M_{1}}{\mu_{0}}\right)^{1}\ldots\left(\frac{M_{N-3}}{\mu_{N-4}}\right)^{1}}} = \\
& = \frac{\prod_{i=0}^{N-1}M_{i}}{\prod_{i=0}^{N-2}M_{i}} = M_{N-1}
\end{aligned}
\end{equation}
where $M_{i}$ is the $i$-th leading principal minor, and $\mu_{0} = M_{0}$.
This means that the only constraint on the coefficient matrix is $M_{N-1}\neq0$.
\end{proof}

\textbf{Remark:} above, we have used the following recurrent formula $M_{i} = \prod_{i=0}^{i}\mu_{i}$.

\textbf{Remark:} this theorem coincides with the theorem we have proven in~\cite{Veneva_2018b}, because no matter what the number of diagonals ($2\times M + 1$) within the coefficient matrix is, the logic remains.

The requirement on the coefficient matrix to be nonsingular is not limiting at
all since this is a standard requirement so as the SLAE to have only one solution.

\subsection{Number of computational steps}
The calculation of $\alpha_{i}^{k},\mu_{i},\alpha_{i}^{M+1},\alpha_{i}^{M+2},\ldots,\alpha_{i}^{2M}$, and $z_{i}$ depends on the results of the calculation of $\alpha_{i-j}^{M+k}$, and $z_{i-j}$. On the other hand, the calculation of $x_{i}$ depends on the results of the calculation of $\alpha_{i}^{M+1}, \alpha_{i}^{M+2},\ldots,\alpha_{i}^{2M}, z_{i}$, and $x_{i+1}, x_{i+2},\ldots,x_{i+M}$. This makes the multi-diagonal numerical method inherently serial. It takes $2\times N$ steps overall, where $N$ is the number of equations in the initial SLAE.

\subsection{Complexity}

The amount of operations per expression are summarized in Table~\ref{tab:complexity}. Thus, the overall complexity of the multi-diagonal numerical algorithms is:
\begin{equation}
2NM^{2} + 5NM + N -\dfrac{4M^{3}}{3} - \dfrac{7M^{2}}{2} - \dfrac{13M}{6},
\end{equation}
where $N$ is the number of rows in the initial coefficient matrix. Hence, the multi-diagonal numerical method requires only $O(N)$ operations (provided that $M<<N$) for finding the solution, and beats the Gaussian elimination which requires $O(N^3)$ operations. 

% \begin{itemize}
% \item 
% $\alpha_{i}^{k}, k = 2,\ldots,M$ is equal to\\
% $(N - 1 - M + 1)\times\sum_{k = 1}^{M}((k-1)\times2) = (N - M)\times(M^{2} - M)$\\ (for $M = 2$ it is $2(N-2)$, for $M = 3$ it is $6(N-3)$, for $M=4$ it is $12(N-4)$).
% % this calculation is made for i >= M
% \item 
% $\mu_{i}, i < M$:\\ 
% $\sum_{k=0}^{M-1}(2k) = M^{2}-M$\\
% (for $M = 2$ it is $2$, for $M = 3$ it is $6$, for $M = 4$ it is $12$).
% \item 
% $\mu_{i}, i\geq M$:\\ 
% $(N - 1 - M + 1)\times 2M$\\ 
% (for $M = 2$ it is $4N - 8$, for $M = 3$ it is $6N - 18$, for $M = 4$ it is $8N-32$).
% % this calculation is made for i >= M
% \item 
% $\alpha_{i}^{M+k}, k = 1,2,\ldots,M$ is equal to\\ 
% $\sum_{k=1}^{M}((N-M-k)\times((M-k)2+1))$ \\
% (for $M = 2$ it is $4N - 13$, for $M = 3$ it is $9N - 41$, for $M = 4$ it is $16N - 94$).
% % this calculation is made for i >= M
% \item 
% $z_{i}, i < M$:\\ 
% $\sum_{k=0}^{M-1}(2k+1) = M^{2}$\\
% (for $M = 2$ it is $4$, for $M = 3$ it is $9$, for $M = 4$ it is $16$).
% \item 
% $z_{i}, i\geq M$:\\ 
% $(N - 1 - M + 1)\times(2M + 1)$\\
% (for $M = 2$ it is $5N - 10$, for $M = 3$ it is $7N - 21$, for $M = 4$ it is $9N-36$).
% \item 
% $x_{N-k}, k = 1,\ldots, M$ is equal to \\
% $\sum_{k = 1}^{M}((k-1)\times2) = (M^{2} - M)$\\
% (for $M = 2$ it is $2$, for $M = 3$ it is $6$, for $M = 4$ it is $12$).
% \item 
% $x_{N-k}, k = M + 1,\ldots, N$ is equal to \\
% $(N - (M+1) + 1)\times 2M$\\
% (for $M = 2$ it is $4N-8$, for $M = 3$ it is $6N - 18$, for $M = 4$ it is $8N-32$).
% \end{itemize}

%\begin{landscape}
\renewcommand{\arraystretch}{2.0}
%\begin{table}[H]
%\centering 
\begin{longtable}{|l|l|r|l|l|l|}
\caption{Complexity per expression for the multi-diagonal numerical algorithm.\label{tab:complexity}}\\\hline
\multirow{2}{*}{expression} & \multirow{2}{*}{\# operations} & \multirow{2}{*}{simplified form of \# ops} & \multicolumn{3}{c|}{examples}\\\cline{4-6}
& & & $M = 2$ & $M = 3$ & $M = 4$ \\\hline
\endhead
$\alpha_{i}^{k}, i < M,$ & \addstackgap{$\displaystyle\sum_{k = 1}^{M-1}(2\times(1+2+\ldots k-1)) = $} & $\dfrac{(M-1)\times M \times (2M-1)}{6}$ & \multirow{2}{*}{$0$} & \multirow{2}{*}{$2$} & \multirow{2}{*}{$8$}\\
$k = 2,\ldots,M$ & \hspace{2em}$\displaystyle\sum_{k = 1}^{M-1}\left(2\times\dfrac{(k-1)\times k}{2}\right)$ & \hspace{2em}$-\dfrac{(M-1)\times M}{2}$ & & & \\[1em]
\hline
$\alpha_{i}^{k}, i \geq M,$ & $(N - 1 - M + 1)$ & \multirow{2}{*}{$(N - M)\times(M^{2} - M)$} & \multirow{2}{*}{$2(N-2)$} & \multirow{2}{*}{$6(N-3)$} & \multirow{2}{*}{$12(N-4)$} \\
$ k = 2,\ldots,M$ & \hspace{2em}$\times\displaystyle\sum_{k = 1}^{M}((k-1)\times2)$ & & & & \\[1em]
\hline
$\mu_{i}, i < M$ & 
\addstackgap{$\displaystyle\sum_{k=0}^{M-1}(2k)$} & $M^{2}-M$ & $2$ & $6$ & $12$
\\\hline
$\mu_{i}, i\geq M$ & 
$(N - 1 - M + 1)\times 2M$ & $2MN - 2M^{2}$ & 
$4N - 8$ & $6N - 18$ & $8N-32$ \\\hline
\multirow{2}{*}{$\alpha_{i}^{M+k}, i < M, $} &  
\addstackgap{$\displaystyle\sum_{k=1}^{M}\left(\displaystyle\sum_{i=0}^{M-k}
(i\times2 + 1)\right.$} & $M^{3}$ & 
\multirow{3}{*}{$6$} & \multirow{3}{*}{$19$} & \multirow{3}{*}{$44$} \\[1em]
$k = 1,2,\ldots,M$ & \hspace{0em}$\left.+(k-1)\times((M-k)\times2+1)\vphantom{\displaystyle\sum_{i=0}^{M-k}}\right)$ & \hspace{0em}$-\dfrac{M\times(M+1)\times(2M+1)}{6}$ & & & \\
&  & $\hspace{0em}+\dfrac{M\times(M+1)}{2}$ & & &
\\[1em]\hline
$\alpha_{i}^{M+k}, i \geq M,$ &  
\addstackgap{$\displaystyle\sum_{k=1}^{M}\left((N-(M+k))\right.$} & $NM^{2}-2M^{3}-M^{2}$ & 
\multirow{3}{*}{$4N - 13$} & \multirow{3}{*}{$9N - 41$} & \multirow{3}{*}{$16N - 94$} \\
$k = 1,2,\ldots,M$ & \hspace{2em}$\left.\times((M-k)\times2+1)\right)$ & \hspace{0em}$+\dfrac{M\times(M+1)\times(2M+1)}{3}$ & & &
\\[0.5em]
& & \hspace{0em}$-\dfrac{M\times(M+1)}{2}$ & & &
\\[1em]\hline
$z_{i}, i < M$  & 
\addstackgap{$\displaystyle\sum_{k=0}^{M-1}(2k+1)$} & $M^{2}$ & $4$ & $9$ & $16$ \\\hline
$z_{i}, i\geq M$ &  
$(N - 1 - M + 1)\times(2M + 1)$ & $2NM + N - M - 2M^{2}$ &
$5N - 10$ & $7N - 21$ & $9N-36$\\\hline\pagebreak\hline
$x_{N-k},$ & 
\multirow{2}{*}{$\displaystyle\sum_{k = 1}^{M}((k-1)\times2)$} & \multirow{2}{*}{$M^{2} - M$} &
\multirow{2}{*}{$2$} & \multirow{2}{*}{$6$} & \multirow{2}{*}{$12$} \\
$k = 1,\ldots, M$ & & & & &\\\hline
$x_{N-k}, $ &
\multirow{2}{*}{$(N - (M+1) + 1)\times 2M$} & \multirow{2}{*}{$2NM-2M^{2}$} &
\multirow{2}{*}{$4N-8$} & \multirow{2}{*}{$6N - 18$} & \multirow{2}{*}{$8N-32$}\\
$k = M + 1,\ldots, N$ & & & & & \\\hline\hline
\multicolumn{1}{|c|}{Total} & \multicolumn{2}{c|}{\addstackgap{$2NM^{2} + 5NM + N -\dfrac{4M^{3}}{3} - \dfrac{7M^{2}}{2} - \dfrac{13M}{6}$}} & $19N-29$ & $34N-74$ & $53N-150$\\\hline
\end{longtable}
%\end{table}

\section{Conclusions}
A generalised symbolic algorithm for solving 
systems of linear algebraic equations with multi-diagonal coefficient matrices was formulated and presented in pseudocode. Some notes on the correctness of the algorithm were made. Formula for the complexity of the multi-diagonal numerical algorithm was obtained.

\section*{Acknowledgments} The author would like to thank Dr.\,Alexander Ayriyan (JINR), and Prof.\,Toshiyuki Imamura (R-CCS, RIKEN) for their precious comments.

%%%%%%%%%%%%%%%%%%%%%%%%%%%%%%%%%%%%%%%%%%%%%%%%%%%%%%%%%%%%%%%%%%%%%%%%%%%%%%%
%\cleardoublepage
\thispagestyle{empty}
\IEEEpeerreviewmaketitle

\end{document}